\documentclass[preprint,review,12pt]{elsarticle}

\usepackage{amssymb}
\usepackage{amsmath}

\journal{Journal of Physics: Condensed Matter}

\begin{document}

\begin{frontmatter}



\title{ CoFeB/MgO/CoFeB structures with orthogonal easy axes: perpendicular anisotropy and damping}


\author[imp]{H.~G\l{}owi\'nski}

\author[acmin]{A.~\.Zywczak}

\author[st]{J.~Wrona}

\author[imp] {A. Krysztofik}

\author[imp] {I. Go\'{s}cia\'{n}ska}

\author[agh,agh2]{T.~Stobiecki}

\author[imp]{J.~Dubowik\corref{cor1}}
\ead{dubowik@ifmpan.poznan.pl}

\cortext[cor1]{Corresponding author}

\address[imp]{Institute of Molecular Physics, Polish Academy of Sciences, ul. Smoluchowskiego 17, 60-179 Poznan, Poland}

\address[acmin]{AGH University of Science and Technology, Academic Centre of Materials and Nanotechnology, Al. Mickiewicza 30, 30-059 Krakow, Poland}

\address[st]{Singulus Technologies AG, Hanauer Landstrasse 103, 63796 Kahl am Main, Germany}

\address[agh]{AGH University of Science and Technology, Department of Electronics, Al. Mickiewicza 30, 30-059 Krakow, Poland}

\address[agh2]{AGH University of Science and Technology, Faculty of Physics and Applied Computer Science, Al. Mickiewicza 30, 30-059, Kraków Poland}

\begin{abstract}
We report on the Gilbert damping parameter $\alpha$, the effective magnetization $4\pi M_{eff}$, and the asymmetry of the  $g$-factor in bottom-CoFeB(0.93~nm)/MgO(0.90--1.25~nm)/CoFeB(1.31~nm)-top  as-deposited systems.
 Magnetization  of CoFeB layers exhibits a specific noncollinear configuration with orthogonal easy axes and with $4\pi M_{eff}$ values of $+2.2$ kG and $-2.3$ kG for the bottom and top layers, respectively.
We show that  $4\pi M_{eff}$ depends on the asymmetry $g_\perp - g_\parallel$ of the  $g$-factor measured in the perpendicular and the in-plane directions revealing  a highly nonlinear relationship. In contrast, the Gilbert damping is practically the same for both layers. Annealing  of the films results in  collinear easy axes perpendicular to the plane  for both layers. However, the linewidth  is strongly increased due to enhanced inhomogeneous broadening.

\end{abstract}

\begin{keyword}
ferromagnetic resonance \sep perpendicular magnetic anisotropy \sep magnetization precession damping
\PACS 75.30.Gw \sep 75.70.Tj \sep 75.78.-n \sep 76.50.+g

\end{keyword}

\end{frontmatter}



\section{Introduction}
\label{Intro}
CoFeB/MgO/CoFeB systems are extensively employed in magnetic tunnel junctions (MTJs), which are  important for modern spintronic devices such as read-heads and magnetic random-access memory \cite{Dieny2017}.
In these applications the two key features are the perpendicular magnetic anisotropy (PMA) with PMA constant $K_{\bot}$ and magnetization damping with inhomogeneous (extrinsic) and Gilbert (intrinsic) contributions to the ferromagnetic resonance  (FMR) linewidth.

The FMR linewidth is usually enhanced in Ta/CoFeB/MgO stacks for which the values of PMA and   the Gilbert damping parameter $\alpha$ are scattered \cite{ikeda2010,devolder2013,Iihama2014}. Recent experimental results \cite{Iihama2014,Sabino} indicate that there is no correlation between $K_{\bot}$ and $\alpha$ in these systems. Specifically,  $\alpha$ is approximately constant while the PMA tends to improve on annealing. However, systems with a high PMA have often an increased linewidth due to an inhomogeneous broadening \cite{Beaujour2009,shaw2014} so that an extrinsic contribution to the linewidth may be as high as 400--500 Oe \cite{sato} despite $\alpha$ is of 0.01 -- 0.02 in these systems. An increase in linewidth is attributed to an angular dispersion of the easy PMA axis, which results in a high inhomogeneous broadening attributed to the zero-frequency linewidth $\Delta H_0$ \cite{Beaujour2009}.

It has been shown that PMA in CoFe/Ni multilayers is linearly proportional to the orbital-moment asymmetry \cite{shaw2014,shaw2013precise} in accordance with the Bruno's model [see Ref.~\cite{shaw2014} for discussion]. On the other hand,  substantial PMA in Ta/CoFeB/MgO systems \cite{ikeda2010} has been considered as related to   an inhomogeneous concentration of the anisotropy at the interface \cite{Sun} so that the Bruno's model   may be not valid in this case. Based on our experimental results,  we aim to shed some light on possible correlation between  asymmetry of the  $g$-factor and the effective magnetization $4\pi M_{eff}$ , which are the magnetic parameters measured directly in a broadband FMR experiment. According to well known Kittel's formula, a departure from the free electron $g$-factor is proportional to $\mu_{L}/\mu_{S}$ \cite{Kittel1949} so that we can discuss  the asymmetry of the $g$-factor as well as on the asymmetry of the orbital moment on equal footing. Here, we prefer to use asymmetry in $g$-factor for evaluating the relationship between  orbital moment and PMA.

As far as we know, FMR has not yet been thoroughly investigated in "full"  Ta/CoFeB/MgO/CoFeB/Ta  MTJ structures. In particular, a dependence of PMA on the asymmetry in the $g$-factor has not yet been proved in CoFeB/MgO/CoFeB systems.
In this paper, we aim to independently characterize each CoFeB layer separated by a MgO tunnel barrier in terms of the $\alpha$ parameter and $4\pi M_{eff}$.  By analyzing FMR measurements in the in-plane and out-of-plane configurations, we find that PMA correlates with the $g$-factor asymmetry in a highly nonlinear relationship.

\section{Experimental methods}
\label{Exp}

The samples were sputtered in an Ar atmosphere using a Singulus Timaris PVD Cluster Tool. The CoFeB magnetic films were deposited by dc-sputtering from a single Co$_{40}$Fe$_{40}$B$_{20}$ target, whereas the MgO barriers were deposited by rf-sputtering directly from a sintered MgO target.
The samples were deposited on an oxidized silicon wafer with 5 Ta/ 20 Ru /Ta 3 buffer layers and capped with 5 Ta/ 5 Ru (numbers indicate the nominal thickness in nanometres).
 The studied structures   consist of two ferromagnetic CoFeB (0.93 nm -- bottom and 1.31 nm -- top) films separated by a MgO barrier of different thicknesses (0.90, 1.1,  and 1.25~nm). It is important to note that we investigated the as-deposited samples so that the CoFeB layers were amorphous \cite{devolder2013,kuswik_cofeb}. The effect of annealing treatment ($330^{o}$C for 1 hr) on magnetic properties of the system will be discussed at the end of the paper.

Hysteresis loops of the samples were measured by vibrating sample magnetometer (VSM) with the perpendicular and in-plane magnetic fields. The saturation magnetization $M_s$ of 1200 G in the as-deposited state was determined from magnetic moment per unit area vs. CoFeB thickness dependencies \cite{frankowski2015}. To investigate anisotropy and damping in studied samples, vector network analyzer ferromagnetic resonance (VNA-FMR) spectra of the $S_{21}$ parameter were analyzed \cite{NembachAnalizaVNA2011}. VNA-FMR was performed at a constant frequency (up to 40~GHz) by sweeping an external magnetic field, which was applied either in-plane or perpendicular to the sample plane. These two configurations will be referred to as the in-plane and out-of-plane configurations.

Experimental data were fitted using the Kittel formula
\begin{equation}
\frac{\omega}{\gamma_{\parallel}}=\sqrt{\left(H_r+H_a\right)\left(H_r+H_a+4\pi M_{eff}\right)}
\label{eq:inplaneKittel}
\end{equation}
for the in-plane configuration
and
 \begin{equation}
\frac{\omega}{\gamma_{\perp}}=\left(H_r-4\pi M_{eff}\right)
\label{eq:perpKittel}
\end{equation}
for the out-of-plane configuration, where $\omega=2\pi f$ is the angular microwave frequency, $H_r$ the resonance field, $\gamma_{\parallel,\perp}=g_{\parallel,\perp} \mu_B /\hbar$ the gyromagnetic ratio, $g_\parallel$ and $g_\perp$ are the spectroscopic $g$-factors for the in-plane and out-of-plane configurations, respectively, $\hbar$ the reduced Planck constant, $\mu_B$ the Bohr magneton, and $H_a$ the in-plane uniaxial anisotropy field.
$4\pi M_{eff}=4\pi M_{s}-H_{\bot}$ is the effective magnetization , where $M_s$ is the saturation magnetization, and $H_{\bot}=2K_{\bot}/M_s$ is the perpendicular anisotropy field and $K_{\bot}$ is the perpendicular anisotropy constant. For the in-plane easy axis $4\pi M_{eff}>0$ whereas for the perpendicular to the plane easy axis   $4\pi M_{eff}<0$. According to Eqs.~\eqref{eq:inplaneKittel} and \eqref{eq:perpKittel}, $4\pi M_{eff} = -2~K_{eff}/M_{s}$,  where $K_{eff}$ is the  effective anisotropy constant defined as $ K_{\bot} - 2\pi M_{s}^2$ \cite{gowtham}.

\section{Results and discussion}
\label{results}

\begin{figure}[h]
 \includegraphics[width=10cm]{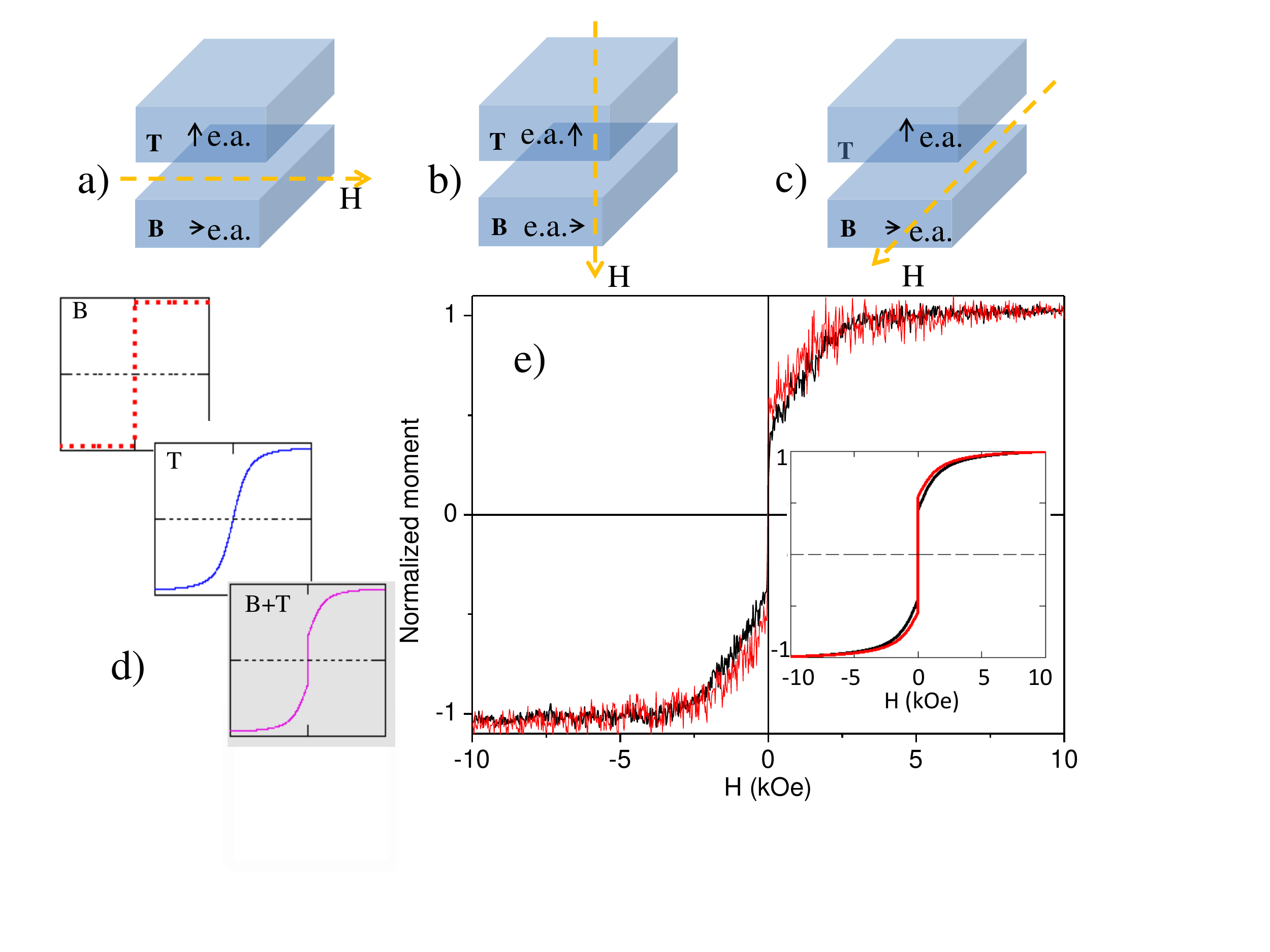}\centering
 \caption{\label{figLoops} (a)-(c) Configurations used for the magnetic measurements with a magnetic field applied perpendicular or parallel to the film plane. (d) Example of schematic pictures of the magnetization reversals of a CoFeB/MgO/CoFeB structure for configuration (a). (e) Hysteresis loops of a CoFeB/MgO/CoFeB structure measured in  configurations (a) - black line and (b) - red line.  The  inset shows schematically the model reversals for configurations (a)-black and (b)-red}.
 \end{figure}

Figure~\ref{figLoops} (e) presents hysteresis loops of the sample with a 1.25~nm thick MgO barrier measured in the out-of-plane (red line) and in-plane configuration (black line).  The shape of the loops in both directions is nearly the same for each configuration as the saturation fields (of $H_{s} \approx 2$ kOe) for both layers have nearly the same magnitude with the opposite signs in $4\pi M_{eff}$. Each hysteresis loop is a sum of  the loops typical for the easy and hard axis and, as explained below, we can infer from magnetization reversals which layer possesses PMA.

Let us assume that the bottom  CoFeB layer (B) has an in-plane easy axis  and the top layer (T) has a perpendicular to the plane easy axis so that their magnetization directions are orthogonal at remanence. Three configurations of a magnetic field $H$ applied for the magnetization measurements are shown in Figs.~\ref{figLoops} (a) - (c). These configurations enable magnetization reversals to be observed with $H$ oriented parallel- (a) (perpendicular- (b)) to the easy axis of B (T) layer, respectively, or perpendicular to both easy axes (c). Further, we will refer to these configurations as (a), (b), and (c) configurations. As it is schematically shown in Fig.~\ref{figLoops} (d), an apparent magnetization reversal of B+T for the configuration (a) is a sum of independent magnetization reversals of B and T. For the perfectly asymmetric structure with $4\pi M_{eff}^{B} = -4\pi M_{eff}^{T}$ with the same thickness (i.e. with the same magnetic moments $M_{S} V^{T,B}$) the apparent magnetization reversals taken in configurations (a) and (b) would overlay. However, as it is seen in Fig.~\ref{figLoops} (e) they do not completely overlay so that the curve taken in the configuration (b) lies a bit higher than that taken in  (a). As it is shown in the inset of (e), a simple model explains that the T layer (i.e. the with nominal thickness $t$ of 1.3 nm) possesses an  easy axis perpendicular to the plane, while the  B layer with $t = 0.93$  nm has an in-plane easy axis.

 In the model, the magnetization reversals in each layer can be approximated with a normalized relation \cite{stearns1994} $M(H,S) = \arctan[H/H_{s}\times \tan(\pi S/2)]/$ $\arctan[H/H_{max}\times \tan(\pi S/2)]$, where $H_{s}$ of 2 kOe is a saturation field for the hard direction and $S$ is defined as a ratio of remanence to the saturation moment. For $H_{\parallel}$ parallel to the easy axis, $S=1$ (B layer in Fig.~\ref{figLoops} (d)) and for $H_{\perp}$ perpendicular to the easy axis (T layer in Fig.~\ref{figLoops} (d)), $S=0.66$ as well as $H_{max}=10$ kOe are arbitrary chosen for the sake of simplicity.  The apparent magnetization curve for configuration (a) is a sum $[t^{B} \times M(H,S=1) + t^{T}\times M(H,S=0.66)]/(t^{B} + t^{T})$. For the configuration (b), $t^{T}$ and $t^{B}$ are reversed in the sum. In order to satisfy the experimental data shown in (e), a ratio $t^{B}/t^{T}=0.79$. It is easily seen that if the B layer had an in-plane easy axis and the T layer had an  easy axis perpendicular to the plane, a curve taken in configuration (b) would lie  lower than that taken in configuration (a).  Hence, the thin B layer is that with the in-plane easy axis.
\begin{figure}[t]
\includegraphics[width=10cm]{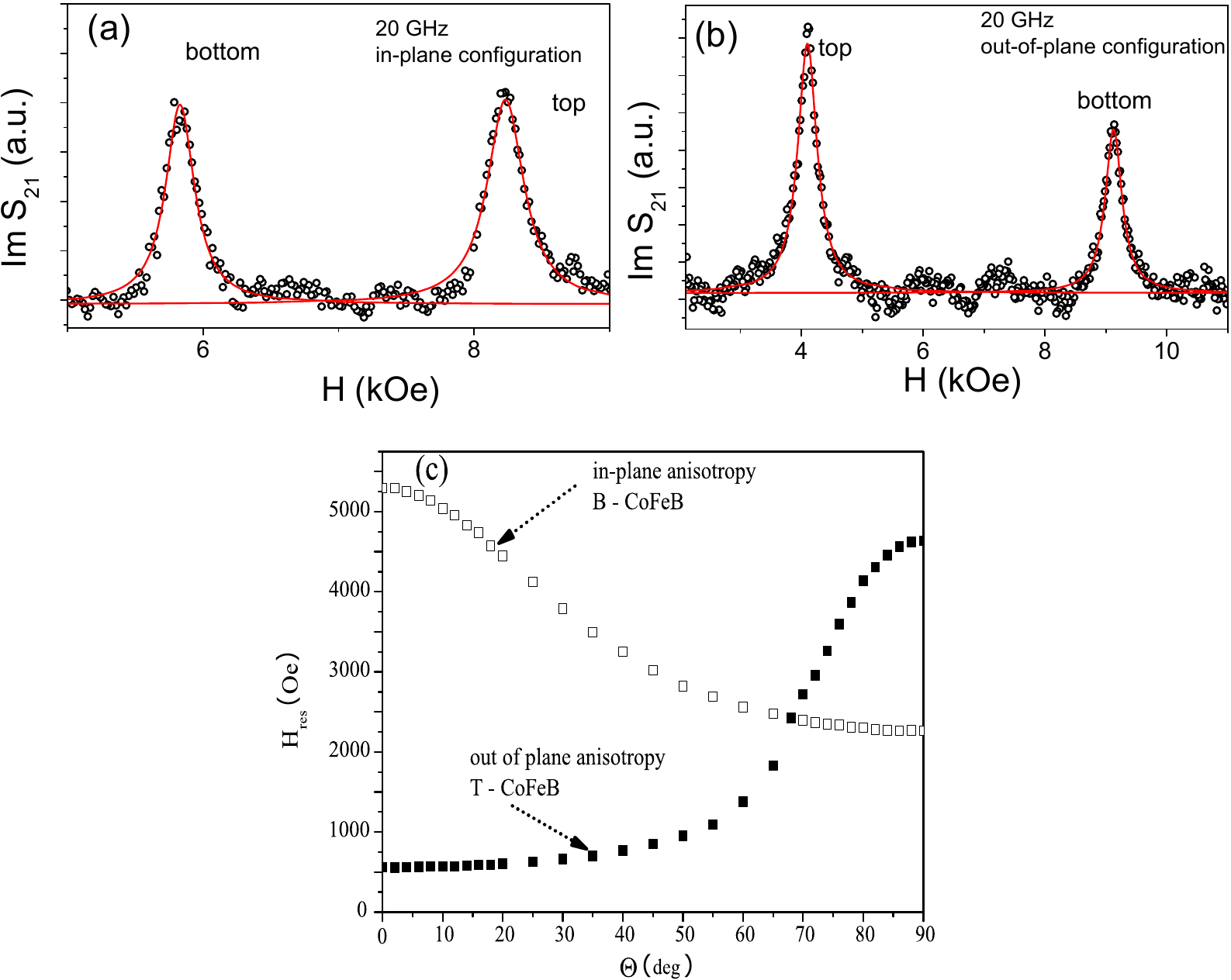}\centering
\caption{\label{figSpectra}Typical VNA-FMR spectrum of the as-deposited CoFeB/MgO(1.25~nm)/CoFeB  structure with resonance peaks from bottom (B) and top (T) layers measured in the in-plane (a) and out-of-plane (b) configurations. Solid red lines represent the Lorentzian fits to the experimental data. (c) Dependence of the FMR field on the polar angle $\Theta$  of applied field in X band (9.1 GHz). The easy axis of magnetization of the B is in the in-plane
orientation. For the T layer, the out-of-plane direction becomes the easy axis.}
\end{figure}

Figures~\ref{figSpectra} (a) and (b) show typical VNA-FMR spectra of the CoFeB/MgO(1.25 nm)/CoFeB system measured (see Figs.~\ref{figLoops}) in configuration (a) and (b) , respectively. Two FMR peaks associated with the bottom and top CoFeB layers are clearly visible. To determine the resonance field $H_r$ and the linewidth $\Delta H$  at constant frequency with a high precision, the spectra were fitted with  Lorentzians   (marked by solid lines in Fig.~\ref{figSpectra} (a) and (b)). Figure~\ref{figSpectra} (c) shows dependencies of the X-band (9.1 GHz) resonance fields of the B and T layers on the polar angle between the film normal and the direction of an applied field. It is clearly seen  that the T layer has $4\pi M_{eff}<0$ (i.e., a perpendicular easy axis) and the B layer with $4\pi M_{eff}>0$ has an in-plane easy axis. From Figs.~\ref{figSpectra} (a) and (b), we can clearly see that the intensity (area under the FMR peak) of the T layer is higher than that of the B layer. This additionally confirms that the bottom layer has the lower magnetic moment than that of the top layer.

\begin{figure}[t]
\includegraphics[height=0.5\textheight]{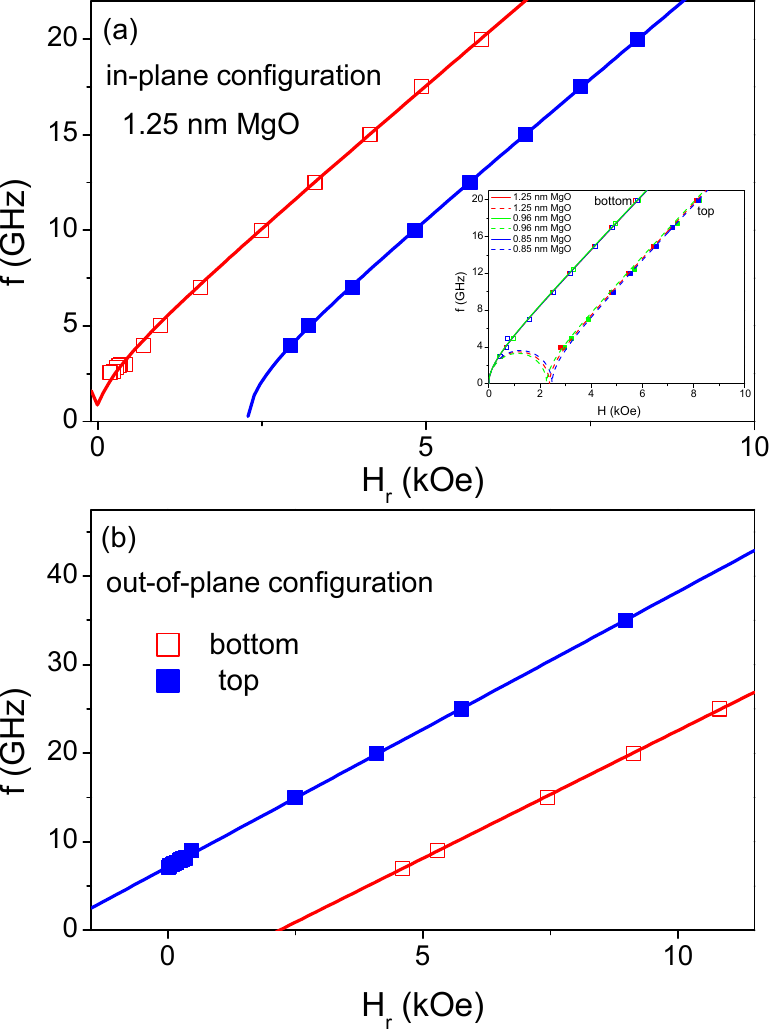}\centering
\caption{\label{figFMRdispersion} FMR dispersion relations of the as-deposited CoFeB/MgO(1.25~nm)/CoFeB structure measured in the in-plane configuration (a) and out-of-plane configuration (b). The solid lines show the fits given in accordance with Eqs.~\eqref{eq:inplaneKittel} and \eqref{eq:perpKittel}. Inset in (a) shows that the fitting parameter practically do not depend on the MgO thickness.}
\end{figure}

 A typical  $H_r$ vs. $f$ dependence, observed for the CoFeB/MgO(1.25~nm)/CoFeB system is shown in Fig.~\ref{figFMRdispersion} (a) and (b) for the in-plane (a) and out-of-plane (b) configuration, respectively. The observed  data points are fitted using Eqs.~\eqref{eq:inplaneKittel} and \eqref{eq:perpKittel}. The values of $4\pi M_{eff}$, obtained from the fitting are found to be of $+ 2.2$~kG and $-2.3$~kG for the bottom and top layers, respectively. The $f$ versus $H_r$ data for the B layer were fitted assuming $H_a$ of 30~Oe as confirmed by VSM  measurements (not shown) in the  configuration presented in Fig.~\ref{figLoops}(c). The values of $g_\parallel$ of the top and bottom layers are equal to 2.04 and 2.08, respectively, in contrast, the values of $g_\perp$ for these layers are 2.06 and 2.22. One can notice the differences in values of $g_\perp$ resulting from clear differences in the slopes of the $f(H_{r})$ dependencies (see, Fig.~\ref{figFMRdispersion} (b)) for the bottom ($\gamma_{\perp}=2.88$ MHz/Oe) and top ($\gamma_{\perp}=3.11$ MHz/Oe) layer, respectively.

  To sum up, VSM and FMR measurements confirmed the presence of orthogonal easy axes in our CoFeB/MgO/CoFeB systems and showed that  the thickness ratio  $t^{B}/t^{T}=0.79$ is slightly higher than the ratio of nominal thickness ($t_{nom}^{B}/t_{nom}^{T}=0.71$). The thinner B layer has an in-plane easy axis while the T layer has a perpendicular easy axis. However, keeping in mind our former studies of a dead magnetic layer (DML) in the Ta/CoFeB/MgO (B) and MgO/CoFeB/Ta (T) structures \cite{frankowski2015} deposited in the same Timaris system, we  estimated DML$^{B}\simeq 0.23$ nm and DML$^{T}\simeq 0.4$. With such asymmetric DMLs the effective thickness $t^{B}_{eff}\simeq 0.7$ nm and $t^{T}_{eff}\simeq 0.9$ nm which satisfies $t^{B}/t^{T}=0.78$.  VNA-FMR measurements, which offer a greater precision than VSM measurements,  give  $4\pi M_{eff}=-2.3$~kG ($K_{\perp}=10.4\times 10^{6}$~erg/cm$^3$) and  $4\pi M_{eff}= +2.2$~kG ($K_{\perp}=7.7 \times 10^{6}$~erg/cm$^3$) for the T and B layers, respectively.
All fitting parameters for a CoFeB/MgO(1.25~nm)/CoFeB structure are juxtaposed in Table~\ref{tab:tabel}. As it is shown in the inset of Fig.~\ref{figFMRdispersion} (a), the thickness of MgO spacer within a range of 0.9 -- 1.25 nm had almost no influence on the fitting parameters, therefore, the values of fitting parameters $4\pi M_{eff}$, $g$, $\alpha$, and $\Delta H_0$ are typical for all samples with various MgO thickness.

\begin{table*}
 \caption{\label{tab:tabel} Parameters determined from VNA-FMR spectra for the as-deposited CoFeB(0.93~nm)/MgO (1.25~nm)/CoFeB(1.31~nm) for the in-plane and out-of-plane configurations: the in-plane anisotropy field ($H_a$), the effective magnetization ($4\pi M_{eff}$), spectroscopic $g$-factors for in-plane and out-of-plane configuration, Gilbert damping ($\alpha$), the frequency-independent FMR linewidth ($\Delta H_0$). The values of the fitting parameters do not depend on the MgO thickness. The values of $g_{\perp}$ are marked by asterisks.}
 \centering
\begin{tabular}{l|ccccc}

 & \multicolumn{5}{c}{In-plane configuration}  \\
 & $H_a$ (Oe) & $4\pi M_{eff}$ (kG) & $g_\parallel$, $g_{\perp}$ & $\alpha$ & $\Delta H_0$ (Oe)\\ \hline
top & 0 & -2.29$\pm$0.05 & 2.04$\pm$0.02 & 0.018$\pm$0.002& 102$\pm$22 \\
bottom & 30 & 2.22$\pm$0.15 & 2.08$\pm$0.03 & 0.017$\pm$0.002 & 69$\pm$23\\
\hline
& \multicolumn{5}{c}{Out-of-plane configuration}  \\
top & -- &-2.3$\pm$0.01 & 2.22$\pm$0.01~$^{\star}$ & 0.018$\pm$0.001& 95$\pm$13\\
bottom & -- & 2.19$\pm$0.04 & 2.06$\pm$0.02~$^{\star}$ & 0.017$\pm$0.003& 160$\pm$30\\

\end{tabular}%
\end{table*}%

Although it is counter-intuitive that the thinner B layer possesses an in-plane easy axis, the same feature has been reported for other Ta/CoFeB(1 nm)/MgO systems deposited in the same Timaris equipment \cite{aleksandrov}.
 Similar effect has been recently observed in a substrate/MgO/CoFeB/Ta/CoFeB/MgO structure, where the thicker CoFeB layer exhibits a strong PMA in contrast to the relatively weak PMA in the thinner CoFeB layer \cite{Shi2017jjap,Shi2017prb}.
  It is possible that the growth mode of the MgO layer in contact with an amorphous CoFeB layer might be responsible. The perpendicular  anisotropy in these systems originates from the CoFe/MgO interface \cite{Yang2011}. The structure of the unannealed CoFeB layers is amorphous regardless of underlying layers, whereas the MgO barrier deposited on the amorphous CoFeB has an amorphous structure of up to four monolayers (that is about 0.9~nm) \cite{Yuasa2005}. Hence, there are subtle differences between the CoFeB/MgO (bottom) and MgO/CoFeB (top) interfaces; the interface of the bottom CoFeB layer is mainly amorphous whereas the interface of the top layer is crystalline, because the  barrier thickness of the investigated samples  is above the transition from amorphous to crystalline phase. Therefore, different structures for the CoFeB/MgO interfaces may result in different values of anisotropy constant. Another  explanation is that the measured dependence $K_{eff} \times t_{eff}$ vs. $t_{eff}$ in films with PMA is often strongly nonlinear due to either intermixing at interfaces \cite{Liu} or magnetoelastic effects \cite{gowtham}, with $K_{eff} \times t_{eff}$ exhibiting a maximum as a function of decreasing $t_{eff}$ and with the PMA eventually being lost for small $t_{eff}$ of, for example, 0.7 nm.

The values of $g$ factor yield the ratio of the orbital $\mu_L$ and spin $\mu_S$ magnetic moments in accordance with equation \cite{shaw2013precise,Kittel1949}
 \begin{equation}
\frac{\mu_L}{\mu_S}=\frac{g-2}{2},
 \label{eq:kittel}
 \end{equation}
where $\mu_S=\mu_B$. Hence, the difference between orbital moments $\Delta \mu_L$ along the easy and hard direction in the in-plane [Fig.~\ref{figLoops} (a)] and out-of-plane [Fig.~\ref{figLoops} (b)] configurations is proportional to $(g_\perp - g_\parallel)$ and reads  $\Delta \mu_L = \mu_B (g_\perp - g_\parallel)/2$. $\Delta \mu_L$  is of 0.09$\mu_B$ and $-0.01 {\mu_B} $ for the T and B layer, respectively.

In CoFe/Ni multilayers \cite{shaw2014},  the PMA has been shown to be proportional to the orbital moment anisotropy in accordance to Bruno model \cite{shaw_2013}. However, in the case of the CoFeB/MgO systems this direct relationship between the orbital moment asymmetry and the perpendicular anisotropy is not fulfilled. As can be seen in Table~\ref{tab:tabel}, $(g_\perp - g_\parallel)\approx 0$ for the B layer corresponds to $4\pi M_{eff} = 2.2$ kG. Hence, while $(g_\perp - g_\parallel)$ is negligible, a decrease in $4\pi M_{eff}$ due to PMA from $4\pi M_{S}=15$ kG to 2.2 kG is substantial. In contrast, $(g_\perp - g_\parallel)\approx 0.18$ is exceptionally large for the T layer, while $4\pi M_{eff}$ merely decreases to - 2.3 kG. In accordance with the earlier report \cite{ShawCoFeB2015}, this confirms that any relationship between the orbital moment asymmetry and the perpendicular anisotropy in CoFeB/MgO systems is highly nonlinear. Of course, other factors controlled by annealing such as disorder at interfaces and over- or underoxidized interfaces would also play a significant role in PMA \cite{Yang2011}. Future work confirming such a nonlinear relationship for a broad range of $t_{CoFeB}$ might resolve this issue.

 At present, there is no doubt that PMA in MgO/CoFeB structures is an interface effect and it is correlated with the presence of oxygen atoms at the interface despite the weak spin-orbit coupling \cite{Yang2011,Li}. The origin of PMA is attributed to hybridization of the O-p with Co(Fe)-d orbitals at the interface \cite{Yang2011} and/or  to a significant contribution of thickness dependent magnetoelastic coupling \cite{gowtham}.   A deviation of the $g$-factor from the 2.0 value is expressed by $g \simeq 2 -4\lambda/ \Delta$ , where $\lambda<0$  is the spin-orbit constant for Fe(Co) and $\Delta$ is the energy levels splitting in the ligand field \cite{Kittel1949}. While the deviation of the $g$-factor is inversely proportional to $\Delta$,  PMA (and hence  $4\pi M_{eff}$) is proportional to the enhanced spin-orbit-induced splitting around the Fermi level \cite{Yang2011}. This may result in a complex relationship between PMA and $g$-factor anisotropy.

\begin{figure}[t]
\includegraphics[height=0.5\textheight]{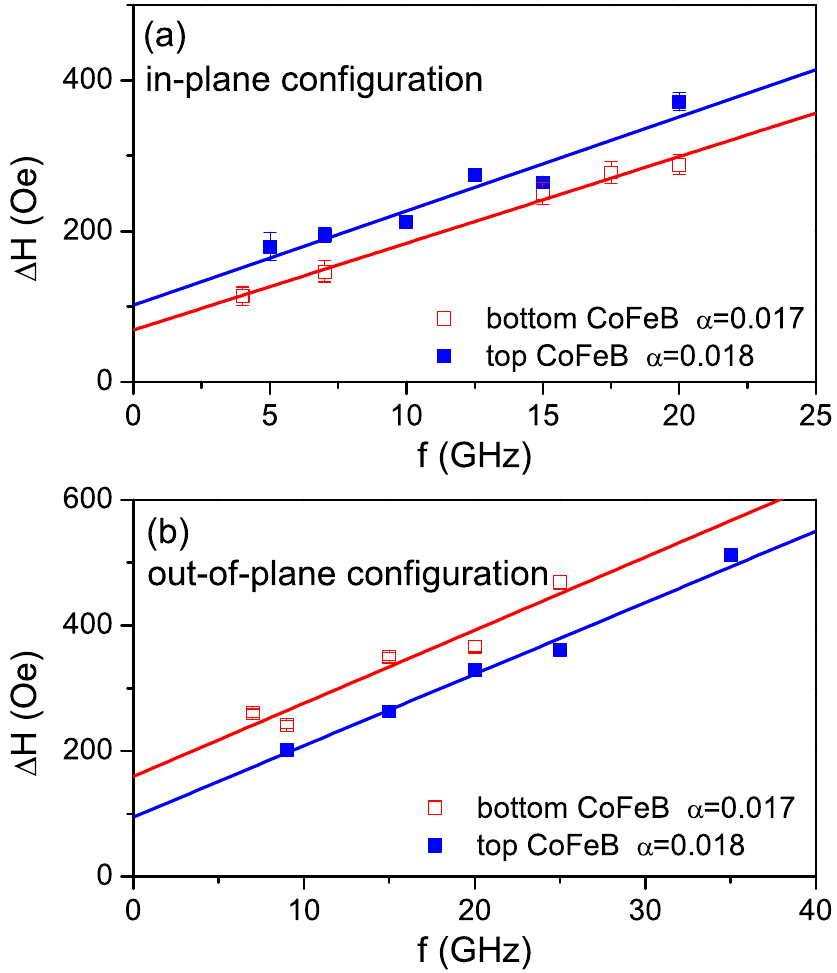}\centering
\caption{\label{figDamp}Linewidth as a function of frequency measured in the in-plane configuration (a) and out-of-plane configuration (b). The $\alpha$ damping parameter is obtained using Eq.~\eqref{eq:Damp}. The thickness of MgO was 1.25~nm.}%
\end{figure}

The Gilbert damping parameter $\alpha$ is evaluated from the dependence of the linewidth  $\Delta H$ on the resonance frequency as shown in Fig.~\ref{figDamp} for the in-plane  (a) and the out-of-plane  (b) configurations. The lines are linear fits to
\begin{equation}
\Delta H=\alpha\frac{4\pi f}{\gamma_{\parallel,\perp}}+\Delta H_0,
 \label{eq:Damp}
 \end{equation}
where $\Delta H_0$ is the inhomogeneous broadening related to CoFeB layer quality. The values of $\alpha$ and $\Delta H_0$ are shown in Table~\ref{tab:tabel}. The top and the bottom layers show almost the same  $\alpha$ of 0.017 - 0.018. This suggests that the damping has no relation to PMA. While $\Delta H_0$ for the top layer is almost the same for both configurations, $\Delta H_0$ for the bottom layer at the (b) configuration is nearly twice as large as that for the (a) configuration.
 Such a behavior suggests that the layer B is rather inhomogeneous with a large angular dispersion of magnetization across the layer \cite{shaw2010,Frankowski201711}.

Spin pumping to Ta layers (which are a part of the buffer and capping layers, as shown in Fig.~\ref{figLoops} (e)) may also influence the damping in CoFeB/MgO/CoFeB systems since magnetization precession induces a spin current to the adjacent nonmagnetic Ta layers that result in an enhanced damping \cite{sato}. This is an interface effect and hence scales inversely proportional to the CoFeB layer thickness. Because the bottom layer with an in-plane easy axis is thinner than the top layer with a perpendicular easy axis, the spin pumping effect affects it more.
To estimate spin pumping effect the standard equation \cite{Tserkovnyak2002} without backflow is used
\begin{equation}
\Delta \alpha= g \mu_B \frac{g_{\downarrow\uparrow}}{4 \pi M_{s} t_{eff}},
\label{spinPump}
\end{equation}
where $t_{eff}$ is the effective thickness of CoFeB and $g_{\downarrow\uparrow}$  is the mixing conductance. The measured damping of both layers is of 0.017 - 0.018, while damping of a bulk CoFeB is around 0.004 \cite{kuswik_cofeb}. Therefore, an increase of $\Delta \alpha$ due to spin pumping is of 0.014 which gives the mixing conductance $g_{\downarrow\uparrow}=0.8$ and $1\times10^{15}$~cm$^{-2}$ for the effective thickness 0.7 nm and 0.9 nm of B and T layer, respectively. The value of mixing conductance  $g_{\downarrow\uparrow}$   for Ta/CoFeB interface found in the literature lies in a broad range from $1.67\times10^{14}$ to $1.4\times10^{15}$~cm$^{-2}$ ~\cite{Allen2015,Cecot2017,Kim20141344,Zhu2017}. Taking into account our simplification (the lack of backflow), this estimation gives the maximal values of mixing conductance. Hence, we can conclude that spin pumping substantially influences the damping in our structures.
It is worth mentioning that the measured $\alpha$ of 0.017 - 0.018 for CoFeB/MgO/CoFeB systems agrees with  $\alpha = 0.015$ for the Ta/CoFeB(1)/MgO structure  reported in \cite{devolder2013}.

\begin{figure}[t]
\includegraphics[height=0.5\textheight]{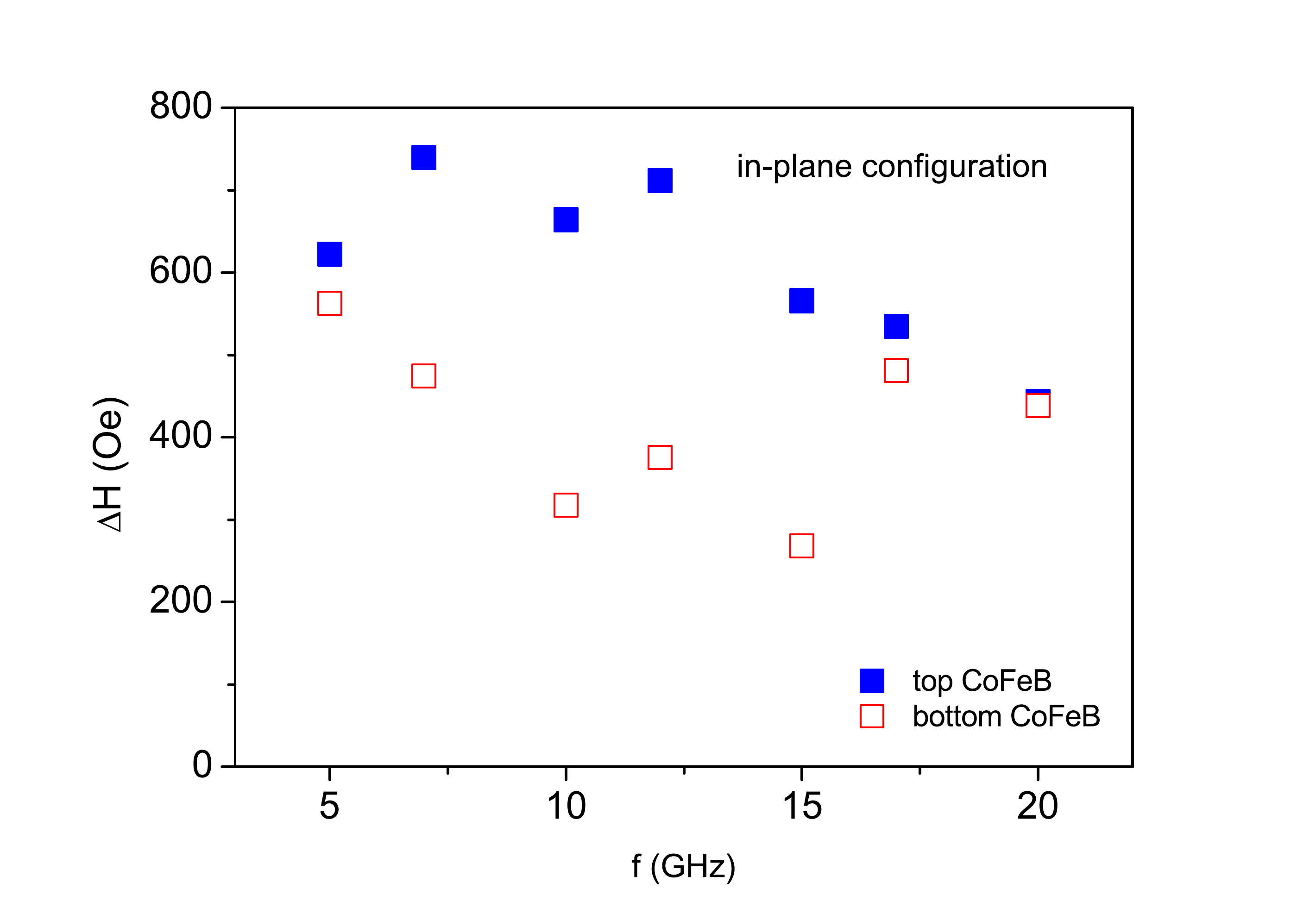}\centering
\caption{\label{fig6linewidth}Linewidth as a function of frequency measured in the in-plane configuration for the annealed structure. The thickness of MgO was 1.25~nm.}%
\end{figure}

\begin{figure}[t]
\includegraphics[height=0.5\textheight]{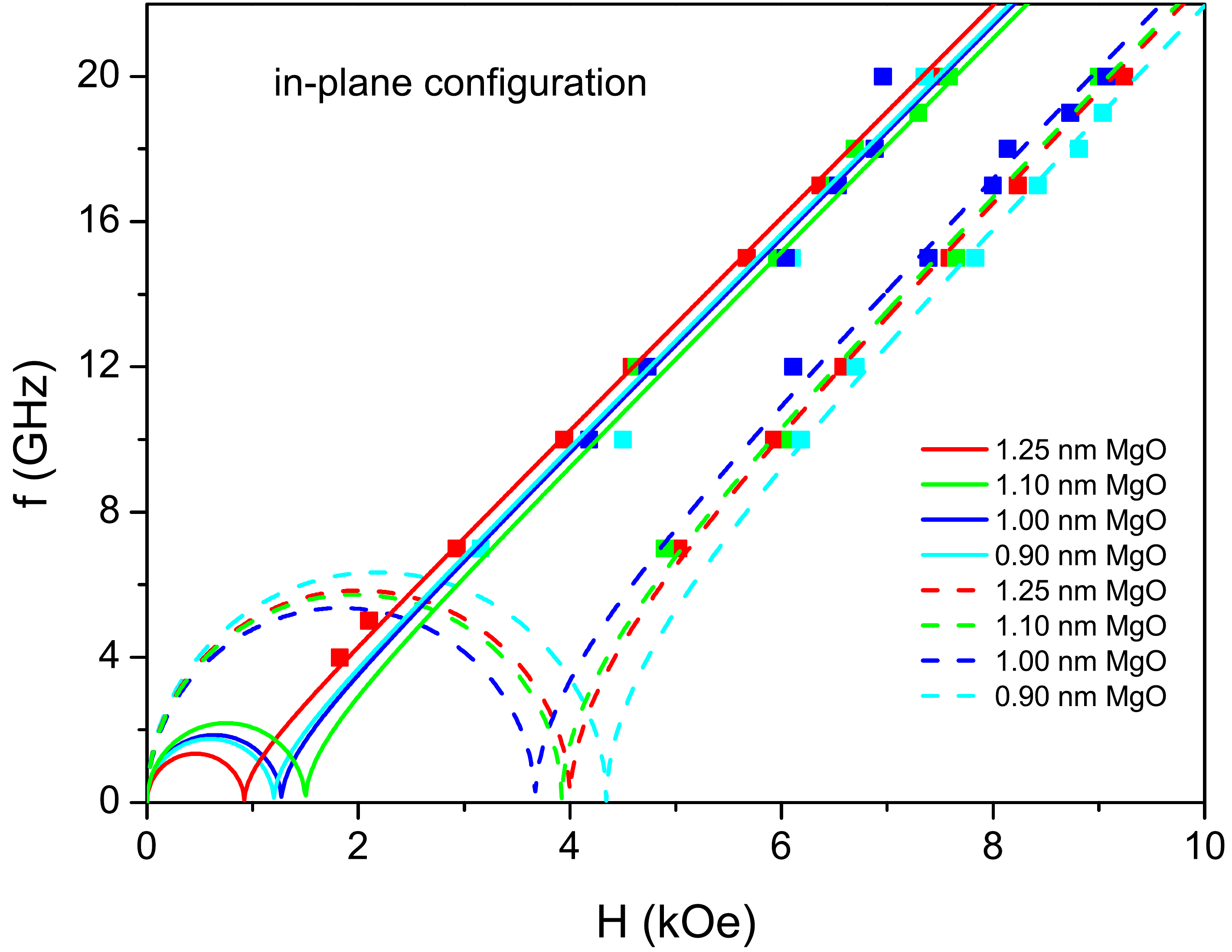}\centering
\caption{\label{Fig7}FMR dispersion relations of CoFeB/MgO(0.9 -- 1.25~nm)/CoFeB annealed structure measured in the in-plane configuration.}%
\end{figure}

Finally, we would like to make a further comment on postdeposition annealing of our CoFeB/MgO/CoFeB systems.  We found that annealing at $330^{o}$C for 1 hr, beside increasing $M_{s}$ to 1500 G, enhances also PMA so that both layers possess easy axes perpendicular to the plane. $4\pi M_{eff}$ attains -1 kG  and -4 kG for the B and T layers, respectively. We found that an increase in $K_{\perp}$ of $7.7 \times 10^{6}$~erg/cm $^3$ equally contributes to both layers and, for example, $K_{\perp} = 17 \times 10^{6}$~erg/cm $^3$ for the T layer. On the other hand, the linewidth $\Delta H$ strongly broadens to $\sim 400$ Oe and $\sim 700$ Oe for the B layer and the T layer, respectively. These values are in agreement with recently reported values for a similar systems \cite{aleksandrov}. Moreover, as it is shown in Fig.~\ref{fig6linewidth}, $\Delta H$ does not follow the linear dependence described by Eq.~\eqref{eq:Damp}. Therefore, it is impossible to determine $\alpha$ precisely for the annealed systems. Such a behavior of $\Delta H$ and the decreased remanence with respect to the saturation magnetization (see, \cite{aleksandrov}) both confirm a strong angular dispersion of the easy PMA axis in both layers.
It has been observed that with increasing PMA the dispersion of anisotropy also increases \cite{Beaujour2009,shaw2014,Frankowski201711}.  As a result, dispersion in PMA leads to a large two magnon scattering contribution to the linewidth for in-plane magnetization and to an enhanced Gilbert damping \cite{Beaujour2009}.
While the magnetic parameters practically do not depend on the MgO thickness in as-deposited structures, the annealed structures show a substantial spread in $4\pi M_{eff}$ as it is shown in Fig.~\ref{Fig7}, which may imply some different CoFeB/MgO interfaces due to, for example, boron diffusion \cite{Cecot2017,Sankha2009}.

\section{Conclusion}
\label{conclusion}
We investigated the CoFeB/MgO/CoFeB  as-deposited systems with the  in-plane  and out-of-plane  orthogonal  easy axes  due to the substantial difference in PMA for the bottom (B) and the  top (T) CoFeB layers, respectively. The T and the B layer  had comparable Gilbert damping $\alpha$ suggesting that there is no correlation between the Gilbert damping and  PMA. We also showed that $4\pi M_{eff}$  correlates with the asymmetry in the $g$-factor (and hence with $\Delta \mu_L$) and this correlation is highly nonlinear. Annealing enhances PMA in both layers but it has detrimental effect on the linewidth, however. Therefore, despite the Gilbert parameter shows no correlation with PMA, it seems that there is some correlation between the linewidth (see Eq.~\ref{eq:Damp}) and PMA in the annealed systems through a combined effect between dispersion of local anisotropy easy axes in crystallites with a high PMA.

\section*{Acknowledgments}
\label{sec:Acknowledgments}

We acknowledge support from the the project ``Marie Sk{\l}odowska-Curie Research and Innovation Staff Exchange (RISE)'' Contract No. 644348 with the European Commission, as part of the Horizon2020 Programme, and partially by the  project NANOSPIN PSPB-045/2010 under a grant from Switzerland through the Swiss Contribution to the enlarged European Union.






\end{document}